# Probability density cloud as a geometrical tool to describe statistics of scattered light

## NATALIA YAITSKOVA

*Hermann-Weinhauser straße 33, 81673 München*
*Corresponding author: nyaitskova@yahoo.com*

**First-order statistics of scattered light is described using the representation of probability density cloud which visualizes a two-dimensional distribution for complex amplitude. The geometric parameters of the cloud are studied in detail and are connected to the statistical properties of phase. The moment-generating function for intensity is obtained in a closed form through these parameters. An example of exponentially modified normal distribution is provided to illustrate the functioning of this geometrical approach. © 2017 Optical Society of America**

http://dx.doi.org/10.1364/AO.99.099999

## 1. INTRODUCTION

In the title of the historical article by Lord Rayleigh: "On the resultant of a large number of vibrations of the same pitch and arbitrary phase" [1], the emphasis is placed on the word "arbitrary", for it is always tempting for a scientist to find a model, an equation, describing the widest range of phenomena. Unfortunately, the larger the range of phenomena the model covers, the heavier it is mathematically, while a compact, ready-for-use expression is often derived for a particular case. The Rayleigh distribution we know is valid only when the phase spans equiprobably the whole range of the possible values from zero to $2\pi$. The generalization, which Lord Rayleigh makes in the last part of his paper, comprises only certain class of phase distributions possessing a $\pi/2$ periodicity. In spite of this limitation, the neat result he has obtained (in his turn being inspired by the book of Émile Verdet [2]) has become the mold for the theory of light statistics for over one century since [3].

The model considered by Lord Rayleigh refers to the so-called strong diffuser. There is another group of random media and reflecting smooth surfaces combined in a model of weak diffuser [4]. In this model the phase is confined within a range smaller than $2\pi$ and the distribution is assumed to be normal (Gaussian). For volume scattering this assumption can be justified: multiple phase shifts along a propagation path permit implementation of the central limit theorem. The ubiquitous use of normal distribution is rather due to the mathematical convenience than to a physical insight: any deviation from the normal distribution brings extensive mathematical difficulties related to complicated expressions involving special functions [5]. Applications often require simplified mathematics; for example, the determination of surface roughness by optical methods relies on the Gaussian model of phase randomness [6].

Nevertheless, extensive experimental evidence shows that different materials and different polishing processes produce various types of surface height distribution. It can be either symmetrical or non-symmetrical, skewed positively or negatively [7, 8]. Some artificially generated biological surfaces have negative skewed height distribution with one or two peaks [9]. In spite of the availability of various statistical distributions, the good old Gaussian model remains the favorite when it comes to analyzing experimental data.

Petr Beckmann [10] first drew his attention to this deficiency and suggested the tool of orthogonal polynomials to model the phenomenon of light scattering from surfaces with arbitrary height distribution. In his approach, the choice of polynomials (Jacobi, Laguerre or Hermite) is however predefined by the type of surface height statistics. The major constraint of scattering theory remains: particular model of height distribution must be postulated prior to any analytical development.

To remove this constraint we suggest a geometrical approach, alternative to the pure analytical method, to describe statistics of the scattered light. We develop a tool of probability cloud: graphical representation of a two-dimensional probability density function (2D-PDF) for complex amplitude. This representation is implicitly utilized by Lord Rayleigh in his paper. The plots of the 2D-PDF are presented in many canonical books on light statistics (for example, in the book by Joseph Goodman [3], paragraph 2.6), but to our knowledge, little attention has been payed to the geometric properties of the cloud: its position, extension, elongation and orientation. A notable exception is the study by Jun Uozumi and Toshimitsu Asakura [11], where this tool is employed to describe on- and off-axis statistics of light in the near-field. The tool of probability cloud provides a cognitive link between stochastic events in the pupil and image planes. On the one side, the geometry of the cloud is related to the distribution of phase and, on the other side, it allows the quantitative prediction of the first order statistics of intensity.

From an arbitrary phase distribution, we derive expressions for the geometric parameters of the cloud: its position, extension, elongation and orientation. The focus of our study is the relation of these parameters to phase statistics. Section 2 describes this relation in the most general way by setting the task into a physical context and then by introducing a mathematical formalism. Section 3 illustrates an application of our approach based on an example of the exponentially modified normal phase distribution. This is only an example, not a restriction of our method. In Section 4 we return to a general description of Section 2 and work with statistics of intensity. We give an expression for the moment-generating function in a closed form through the geometric parameters of the 2D-PDF. As the main subject of the paper is the cloud itself, the material of Section 4 is condensed.

Our study strives to obtain the results for truly arbitrary phase distribution. The only "Gaussian" limitation we preserve is the assumption of a large number of contributors, which means that we deal with a Gaussian complex random variable. Non-Gaussian complex random variable and the theory of non-Gaussian speckles were discussed in our earlier work [12].

## 2. PROBABILITY CLOUD

### A. From a diffraction integral to a random phasor sum

We start by setting our task in the context of diffractive optics. Consider a monochromatic optical wave with a distorted wavefront $\varphi(r)$. The complex amplitude in the pupil plane of an imaging system is given by

$$F(r) = P(r)e^{j\varphi(r)}, \qquad (1)$$

where $r$ is a position in the pupil plane and $P(r)$ is a pupil function, which is one inside the pupil and zero outside it. The amplitude of the wave is constant everywhere within the pupil. Phase randomness might be caused by a turbulent media through which the wave has propagated or by a surface from which the light has reflected. In case of free-space propagation, Eq.(1) describes the complex field directly after the surface and $P(r)$ – the shape of an illumination spot. The imaging system might also introduce phase distortions of its own, like the multi-segmented primary mirror of a giant optical telescope [13].

The complex amplitude in the focal plane of the imaging system (coordinate $w$) is given by the Fourier transform:

$$U(w) = \frac{1}{\lambda f} \iint F(r) \, e^{-j\frac{2\pi}{\lambda f}rw} \, d^2r, \qquad (2)$$

where $\lambda$ is the wavelength and $f$ is the focal distance. The same expression is valid for the complex amplitude in the far-field if the wave reflects from a surface and freely propagates assuming that scattering angles are small, i.e. in the paraxial approximation. Then the focal distance $f$ is replaced by a propagation distance $z$.

The intensity in the image plane $|U(w)|^2$ is a random field, which statistical properties depend on the following factors:

1. First-order statistics of the wavefront distortions $\varphi(r)$ including spatial stationarity or non-stationarity;

2. The ratio between the size of the pupil (or of the illumination spot) and the correlation radius of the complex field $e^{j\varphi(r)}$;

3. Observation point (coordinate $w$).

Further conditions must be set. First, the observation point is chosen to be on-axis, so the diffraction integral becomes

$$U(0) = \frac{1}{\lambda f} \iint F(r) d^2r, \qquad (3)$$

Second, it is assumed that the correlation radius of the complex field $e^{j\varphi(r)}$ is smaller than the size of the pupil so that the pupil includes many coherent cells. Third, it is assumed that the random field $\varphi(r)$ is isotropic and spatially stationary, so that its statistical properties depend neither on direction nor on coordinate in the pupil. Fourth, the illumination is regarded to be uniform: $P(r)$ is constant inside the pupil (within the illumination spot) and zero outside. These assumptions allow splitting the pupil into $n$ identical zones and the diffraction integral after some normalization turns into a random phasor sum with $n$ phasors:

$$A = \frac{1}{\sqrt{n}} \sum_{i=1}^{n} e^{j\varphi_i}. \qquad (4)$$

The random values $\varphi_i$ are identically distributed which results from the assumption of spatial stationarity. They are also mutually uncorrelated which is ensured by the size of the zone. The size of the zone defines also the number of phasors – an important parameter in the study. The definition of the zone and therefore the determination of the number of phasors involves second-order statistics of phase: its autocorrelation function. Some authors prefer working directly with the autocorrelation function, assuming a Gaussian-correlated phase [11, 14]. Nevertheless, this approach does not remove the mathematical complexity of the expressions involved. We prefer to work with the random phasor sum, keeping the number of phasors $n$ as a free parameter. If the source of randomness is telescope segmentation the issue of computing $n$ is trivial (it equals to the number of segments).

### B. Two-dimensional probability density function for the complex amplitude

In the theory of diffused light various parameters and phenomena are said to be Gaussian: Gaussian speckles, Gaussian distribution of phase, Gaussian autocorrelation function of phase and even Gaussian beam. Explicit modelling of the beam shape and of the autocorrelation function can be avoided by introducing the notion of phasors. Although we have assumed a uniform illumination, Eq.(4) can be modified taking into account a non-uniform beam shape. However, we must clarify the difference between Gaussian speckles and Gaussian phases. Although in many studies both assumptions are often made together, they are two different issues. The former concerns the number of phasors: if $n$ is large enough, the central limit theorem can be applied to the complex random variable $A$ and speckles formed by the resulting intensity are called Gaussian. The latter refers to the shape of the probability density function of phase $\varphi_i$ (P-PDF). The number of phasors is defined by the pupil size and the correlation radius of the complex field, i.e. by the second order statistics. The P-PDF is the first order statistics. Here we adopt the first assumption of a large number of phasors but allow the P-PDF to be arbitrary.

In this Subsection we expound the basic mathematical facts about two-dimensional random Gaussian values, conic sections in general and ellipses in particular. If the reader finds this part trivial, he (she) can skip it and go directly to the Subsection 2C.

The assumption of a large number of phasors allows for employing of the central limit theorem and for writing down an expression of the joint probability density function for the real ($R$=Re $A$) and imaginary ($I$=Im $A$) parts of the resultant complex amplitude $A$. From now on we refer to this function as 2D-PDF. It includes the following statistical moments of $R$ and $I$ (we use $\langle \ldots \rangle$ to denote statistical averaging):

$$\begin{aligned} E_R &= \langle R \rangle, \; E_I = \langle I \rangle, \\ \sigma_R^2 &= \langle R^2 \rangle - \langle R \rangle^2, \\ \sigma_I^2 &= \langle I^2 \rangle - \langle I \rangle^2, \\ C_{RI} &= \langle (R - E_R)(I - E_I) \rangle. \end{aligned} \qquad (5)$$

These quantities form the covariance matrix:

$$\hat{c} = \begin{bmatrix} \sigma_R^2 & C_{RI} \\ C_{RI} & \sigma_I^2 \end{bmatrix} \quad (6)$$

with the determinant

$$D_c = \sigma_R^2 \sigma_I^2 - C_{RI}^2. \quad (7)$$

Now we write the 2D-PDF through the covariance matrix:

$$p_A(R,I) = \frac{1}{2\pi\sqrt{D_c}} \exp\left(-\frac{1}{2} \boldsymbol{u}^T \hat{c}^{-1} \boldsymbol{u}\right), \quad (8)$$

where $\boldsymbol{u}^T = \{R - E_R, I - E_I\}$ and $\hat{c}^{-1}$ is the inverse covariance matrix. The product $\boldsymbol{u}^T \hat{c}^{-1} \boldsymbol{u}$ can be written as

$$\boldsymbol{u}^T \hat{c}^{-1} \boldsymbol{u} = \frac{1}{D_c} \boldsymbol{u}^T \hat{s} \boldsymbol{u}, \quad (9)$$

where

$$\hat{s} = \begin{bmatrix} \sigma_I^2 & -C_{RI} \\ -C_{RI} & \sigma_R^2 \end{bmatrix}. \quad (10)$$

The determinant of the matrix $\hat{s}$ also equals $D_c$. The quantity

$$\boldsymbol{u}^T \hat{s} \boldsymbol{u} \stackrel{\text{def}}{=} Q(R,I) = \quad (11)$$

$$\sigma_I^2 (R - E_R)^2 + \sigma_R^2 (I - E_I)^2 - 2C_{RI}(R - E_R)(I - E_I)$$

is a second-order polynomial, and the line $Q(R,I) = const$ is a conic section. Therefore the contour of constant probability density, $p_A(R,I) = const$, is also a conic section.

The conic section can be classifies either by the sign of the discriminant of the quadratic part of $Q(R,I)$ or by the sign of the determinant of the matrix $\hat{s}$. These two quantities differ only by the factor -4. We use the determinant of the matrix $\hat{s}$, which as we have seen coincides with the determinant $D_c$ of the covariance matrix from Eq.(7). Due to the Schwarz's inequality the determinant is non-negative. We first consider the case when $D_c > 0$. The contours of constant probability density are ellipses or circles. We select one contour satisfying the equation

$$Q(R,I) = D_c, \quad (12)$$

and associate the geometric properties of the 2D-PDF with the properties of this particular ellipse. The properties are: position of the center $(\rho, \theta_\rho)$, lengths of semi-major and semi-minor axis $(\sigma_x, \sigma_y)$ and the rotation angle $\Theta$ (Figure 1).

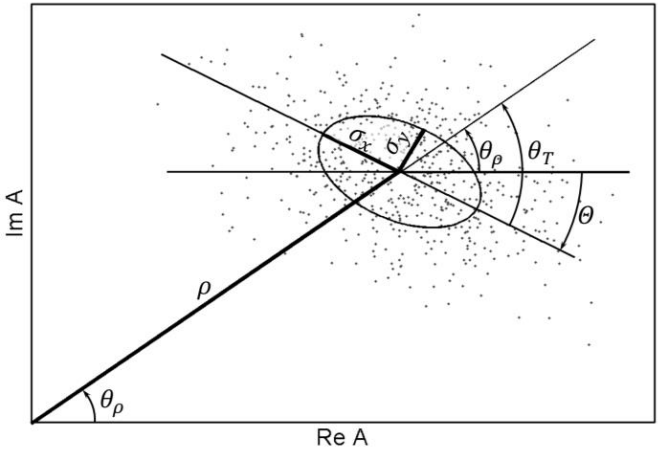

Fig. 1. Two-dimensional probability density cloud and its geometric parameters: distance to the center, semiaxes and the angles. The semi-major axis is always $\sigma_x$ regardless of the orientation of the ellipse.

Consider a new system of coordinates with the origin in the center of the ellipse and the axis coinciding with its axis (Figure 2). Through translation and rotation of the axes, a coordinate $(R,I)$ transforms into a coordinate $(x,y)$ and the equation of the conic section assumes the canonic form. The 2D-PDF becomes:

$$p_A(x,y) = \frac{1}{2\pi\sqrt{D_c'}} \exp\left(-\frac{1}{2D_c'} \boldsymbol{v}^T \widehat{s'} \boldsymbol{v}\right), \quad (13)$$

where $\boldsymbol{v}^T = \{x,y\}$. Matrix

$$\widehat{s'} = \begin{bmatrix} \sigma_y^2 & 0 \\ 0 & \sigma_x^2 \end{bmatrix} \quad (14)$$

is obtained from the matrix $\hat{s}$ by translation and rotation. It has determinant $D_c' = \sigma_x^2 \sigma_y^2$. Two quantities are invariant to the change of the system of coordinates: the trace of the matrix and its determinant. The trace

$$\Sigma = \sigma_R^2 + \sigma_I^2 = \sigma_x^2 + \sigma_y^2 \quad (15)$$

can be regarded as the total size of the ellipse: spread of the 2D-PDF. The discriminant is also preserved: $D_c' = D_c$. However, we prefer to work with another invariant: the eccentricity. We use the definition of eccentricity slightly different from the usual one:

$$e = \frac{\sigma_x^2 - \sigma_y^2}{\sigma_x^2 + \sigma_y^2}. \quad (16)$$

It gives the degree of elongation and it can be expressed through the total size and the determinant as

$$e = \sqrt{1 - \frac{4D_c}{\Sigma^2}}. \quad (17)$$

Solving Eqs. (15) and (16) with respect to $\sigma_x^2$ and $\sigma_y^2$ yields in

$$\sigma_x^2 = \frac{\Sigma}{2}(1+e),$$
$$\sigma_y^2 = \frac{\Sigma}{2}(1-e). \quad (18)$$

The eccentricity changes between zero and one. It equals zero when the semi-axes are identical: $\sigma_x = \sigma_y = \sqrt{\Sigma/2}$. The cloud is then circular, but it does not necessary mean that it is centered at the center of coordinates. In general, position and extension of the cloud do not depend on the eccentricity. When the cloud is circular, the determinant is maximal: $D_c = \Sigma^2/4$.

Consider now the case when $D_c = 0$. The argument of the exponential function in Eq.(8) is a negative infinitely large value everywhere except for the locus $\boldsymbol{u}^T \hat{s} \boldsymbol{u} = Q(R,I) = 0$. The probability density function is zero everywhere except on this locus, where it is infinite. The locus satisfies the equation

$$Q(R,I) = [\sigma_I(R - E_R) - \sigma_R(I - E_I)]^2 = 0 \quad (19)$$

and therefore represents a degenerated conic. If at least one of $\sigma_R$ and $\sigma_I$ is non-zero ($\Sigma \neq 0$), the degenerated conic is a line; if $\sigma_R = \sigma_I = 0$, ($\Sigma = 0$), it is a point at the coordinate $\{R = E_R, I = E_I\}$. For the both types of degenerated conic the eccentricity equals one.

Rotation and translation brings the ellipse to the canonic form, therefore one can find the expressions, relating the coefficients of the covariance matrix and the geometric parameters of the conic section. We present them without the derivation:

$$\sigma_R^2 = \frac{\Sigma}{2}(1 + e\cos\Theta),$$
$$\sigma_I^2 = \frac{\Sigma}{2}(1 - e\cos\Theta),$$
$$C_{RI} = \frac{e\Sigma}{2}\sin\Theta, \quad (20)$$

where $\Theta$ is the rotation angle.

The geometry of the 2D-PDF is fully described if the following five parameters are known: $\rho, \theta_\rho, \Sigma, e$ and $\Theta$. In the next sub-section we relate them to the statistical properties of the phase.

### C. Statistics of phase and geometry of the cloud

The principal input is the characteristic function of phase $\boldsymbol{M}_\varphi(k)$. It is a statistical average of the complex value $e^{jk\varphi}$ and it is related to the P-PDF by the Fourier transform:

$$\boldsymbol{M}_\varphi(k) = \langle e^{jk\varphi} \rangle = \int_0^{2\pi} p(\varphi) e^{jk\varphi} \, d\varphi. \tag{21}$$

Two values of the characteristic function, $\boldsymbol{M}_\varphi(1)$ and $\boldsymbol{M}_\varphi(2)$, give the mean and the variance of the random complex variable $e^{j\varphi}$:

$$\boldsymbol{V} = \boldsymbol{M}_\varphi(1), \boldsymbol{W} = \boldsymbol{M}_\varphi(2) - \boldsymbol{M}_\varphi^2(1). \tag{22}$$

The entities of Eq.(22) are complex values and we represent them through the moduli and the angles:

$$\boldsymbol{V} = V e^{j\theta_V}, \boldsymbol{W} = W e^{j\theta_W}. \tag{23}$$

To relate them to the moments from Eq.(5) we use the equations (2-47) from the Ref. (3) and the properties of complex variable:

$$\begin{aligned}
\boldsymbol{M}_\varphi(1) + \boldsymbol{M}_\varphi(-1) &= \boldsymbol{M}_\varphi(1) + \boldsymbol{M}_\varphi^*(1) \\
&= 2\mathrm{Re}[\boldsymbol{M}_\varphi(1)] = 2V \cos\theta_V, \\
\boldsymbol{M}_\varphi(2) + \boldsymbol{M}_\varphi(-2) - \boldsymbol{M}_\varphi^2(1) - \boldsymbol{M}_\varphi^2(-1) \\
&= 2\mathrm{Re}[\boldsymbol{M}_\varphi(2) - \boldsymbol{M}_\varphi^2(1)] = 2W \cos\theta_W, \\
\boldsymbol{M}_\varphi(2) - \boldsymbol{M}_\varphi(-2) - \boldsymbol{M}_\varphi^2(1) + \boldsymbol{M}_\varphi^2(-1) \\
&= 2j\mathrm{Im}[\boldsymbol{M}_\varphi(2) - \boldsymbol{M}_\varphi^2(1)] = 2jW \sin\theta_W, \\
\boldsymbol{M}_\varphi(1)\boldsymbol{M}_\varphi(-1) &= V^2.
\end{aligned} \tag{24}$$

The moments become

$$\begin{aligned}
E_R &= \sqrt{n} V \cos\theta_V & \text{(a)} \\
E_I &= \sqrt{n} V \sin\theta_V, & \text{(b)} \\
\sigma_R^2 &= \tfrac{1}{2}(1 - V^2 + W \cos\theta_W), & \text{(c)} \\
\sigma_I^2 &= \tfrac{1}{2}(1 - V^2 - W \cos\theta_W), & \text{(d)} \\
C_{RI} &= \tfrac{1}{2} W \sin\theta_W. & \text{(e)}
\end{aligned} \tag{25}$$

Note, that the structure of Eq.(25) can be generalized beyond the model of random phasor sum. Expressions for the moments, derived directly from the diffraction integrals [11, 15], have the same structure, although $\boldsymbol{V}$ and $\boldsymbol{W}$ have a physical meaning other the mean and the variance of $e^{j\varphi}$. In this sense, the results of the next sub-section are not limited to the framework of random phasor sum model.

We relate the parameters of the ellipse to $\boldsymbol{V}$ and $\boldsymbol{W}$. As the center of the ellipse is located at $\{R = E_R, I = E_I\}$ in the $(R, I)$-system of coordinates, the first two parameters come directly from Eq.(25 a, b):

$$\begin{aligned}
\rho &= \sqrt{n} V, \\
\theta_\rho &= \theta_V.
\end{aligned} \tag{26}$$

The total size is obtain by summing Eq.(25 c) and Eq.(25 d):

$$\Sigma = \sigma_R^2 + \sigma_I^2 = 1 - V^2, \tag{27}$$

from where we conclude that the distance to the center of the ellipse and its total size are related:

$$\rho = \sqrt{n(1 - \Sigma)}. \tag{28}$$

The total size varies between zero and one. If the number of phasors is preserved, the cloud shrinks while moving away from the center of the coordinate system. At the farthest distance equal to $\sqrt{n}$ the total size is zero and the cloud shrinks into a point. When the center of the cloud coincides with the center of the system of coordinates, the total size equals one and the cloud is at its maximal spread.

To compute the determinant we combine Eq.(25 c, d, e):

$$D_c = \sigma_R^2 \sigma_I^2 - C_{RI}^2 = \tfrac{1}{4}[(1 - V^2)^2 - W^2] = \tfrac{1}{4}(\Sigma^2 - W^2). \tag{29}$$

Substituting this expression into Eq.(17) for the eccentricity we obtain

$$e = \frac{W}{1 - V^2} = \frac{W}{\Sigma}. \tag{30}$$

Comparing Eq.(25) with Eq.(20) we conclude that

$$\Theta = \frac{\theta_W}{2}. \tag{31}$$

For the application another angle is the more important: the angle between the major axis and the line passing through the center of the ellipse and the center of the coordinate system (Figure 1). We refer to it as inclination angle $\theta_T$. The inclination angle is the sum

$$\theta_T = \theta_\rho + \Theta = \theta_V + \frac{\theta_W}{2}. \tag{32}$$

As it is indicated in Figure 1, we assume that the angles $\theta_\rho$ ($\theta_V$) and $\theta_T$ increase anticlockwise, but $\theta_W$ and $\Theta$ – clockwise.

Eqs. (26), (27), (30) and (31) relate the five parameters of the cloud: $\rho, \theta_\rho, \Sigma, e$ and $\Theta$ to the mean and the variance of the random complex variable $e^{j\varphi}$. In the following, we demonstrate how this approach can be used for a given P-PDF, but before we return shortly to the canonical equation of ellipse.

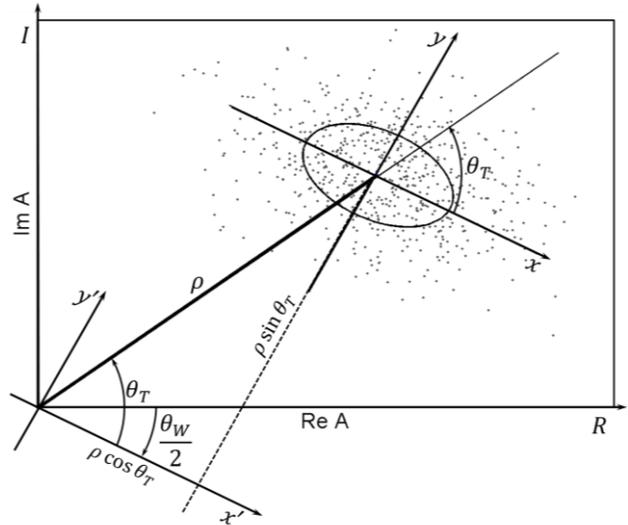

Fig. 2. Two-dimensional probability density cloud in different systems of coordinates: $(R, I)$ – initial system of coordinates, $(x', y')$ - rotated by the angle $\theta_T$, $(x, y)$ – related to the ellipse center and the semi-axes.

To conclude this Section we present equation of the ellipse in the canonic form through the geometric parameters. Consider a system of coordinates $(x', y')$ which is obtained from $(R, I)$-system of coordinates by rotation by the angle $\Theta = \theta_W/2$ (Figure 2). Its center coincides with the center of the system $(R, I)$ and the axes are parallel to the axes of the system $(x,y)$, i. e. are parallel to the axes of the ellipse. The coordinates transformation is

$$x' = R\cos\frac{\theta_W}{2} - I\sin\frac{\theta_W}{2},$$
$$y' = R\sin\frac{\theta_W}{2} + I\cos\frac{\theta_W}{2}. \quad (33)$$

Introducing the polar coordinates $(R = A\cos\theta_A, I = A\sin\theta_A)$ and performing the trigonometric combining, we obtain:

$$x' = A\cos\theta',$$
$$y' = A\sin\theta', \quad (34)$$

where $\theta' = \theta_A + \theta_W/2$. On the other hand, the system of coordinates $(x', y')$ relates to the system of coordinate $(x, y)$ by a parallel shift:

$$x = x' - \rho\cos\theta_T,$$
$$y = y' - \rho\sin\theta_T. \quad (35)$$

The equation of the ellipse becomes

$$\frac{2(A\cos\theta' - \rho\cos\theta_T)^2}{\Sigma(1+e)} + \frac{2(A\sin\theta' - \rho\sin\theta_T)^2}{\Sigma(1-e)} = 1. \quad (36)$$

The inclination angle $\theta_T$ is given by Eq.(32) and $\rho$ is given by Eq. (28). Eq.(36) describes the same curve as Eq. (12), but written in polar coordinates through the geometric parameters of the cloud. We will come back to this result in Section 4.

## 3. EXAMPLE

### A. Exponentially modified normal P-PDF

Suppose that phase distortions are governed by two independent random processes: the first process is normally distributed with zero mean and standard deviation $\sigma \geq 0$:

$$p_1(\varphi) = \frac{1}{\sqrt{2\pi\sigma^2}}\exp\left(-\frac{\varphi^2}{2\sigma^2}\right), \quad (37)$$

and the second process is exponentially distributed with $\tau \geq 0$:

$$p_2(\varphi) = \begin{cases} \frac{1}{\tau}\exp\left(-\frac{\varepsilon\varphi}{\tau}\right), \varepsilon\varphi \geq 0, \\ 0, \varepsilon\varphi < 0, \end{cases} \quad (38)$$

$\varepsilon = \pm 1$. The resultant phase is a sum of these two processes and follows exponentially modified normal distribution law. From the basic theory of random process, we know that the residual probability density function is a convolution of two initial probability density functions. Although for our purpose it is not necessary to know the P-PDF itself, for the completeness we write it down. The exponentially modified normal P-PDF has the following shape:

$$p(\varphi) = \frac{1}{2\tau}\exp\left[\frac{1}{2\tau}\left(\frac{\sigma^2}{\tau} - 2\varepsilon\varphi\right)\right]\text{erfc}\left[\frac{1}{\sqrt{2\sigma^2}}\left(\frac{\sigma^2}{\tau} - \varepsilon\varphi\right)\right], \quad (39)$$

where $\text{erfc}(...)$ is complementary error function and parameter $\varepsilon$ determines the direction of the modification. When $\varepsilon = 1$ the P-PDF is positively skewed, when $\varepsilon = -1$ the P-PDF is negatively skewed.

The first three central moments of the phase are

$$\langle\varphi\rangle = \varepsilon\tau,$$
$$\langle(\varphi - \langle\varphi\rangle)^2\rangle = \sigma^2 + \tau^2, \quad (40)$$
$$\langle(\varphi - \langle\varphi\rangle)^3\rangle = 2\varepsilon\tau^3.$$

The zero mean of the initial normal distribution is shifted by the exponential process in the direction defined by $\varepsilon$. The variance is the sum of the two variances as we adopted the assumption of independent processes. The third central moment included in skewness depends only on $\tau$ and its sign is determined by $\varepsilon$.

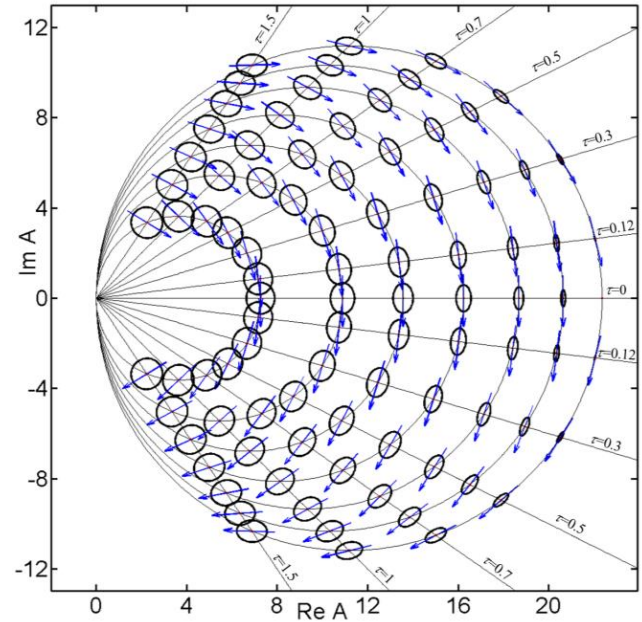

Fig. 3. Family of ellipses showing the geometry of the 2D-PDF for exponentially modified normal phase distribution. Each ellipse corresponds to certain combination $(\tau, \sigma)$. Rays are the lines of constant $\tau = 0, 0.12, 0.3, 0.5, 0.7, 1, 1.5$. Circles are the lines of constant $\sigma = 0, 0.4, 0.6, 0.8, 1, 1.2, 1.5$ (from the outer circle inwards). Upper half-plane: positively skewed P-PDF ($\varepsilon = 1$), lower half-plane: negatively skewed P-PDF ($\varepsilon = -1$). Arrows indicate orientation of ellipses. The angle between an arrow and a ray is inclination angle. The number of phasor $n = 500$.

To describe the geometric properties of the cloud we need to know only two values of the characteristic function. Expression for a characteristic function is often much easier to obtain than expression for a PDF because the characteristic function of a sum of independent random processes is the product of the characteristic functions. For the exponentially modified normal distribution the characteristic function is:

$$M_\varphi(k) = (1 - jk\varepsilon\tau)^{-1}\exp\left(-\frac{k^2\sigma^2}{2}\right). \quad (41)$$

The complex values we need to calculate are

$$V = (1 - j\varepsilon\tau)^{-1}\exp\left(-\frac{\sigma^2}{2}\right), \quad (42)$$
$$W = (1 - j2\varepsilon\tau)^{-1}\exp(-2\sigma^2) - (1 - j\varepsilon\tau)^{-2}\exp(-\sigma^2).$$

In order to extract the parameters of the cloud some algebraic manipulations with complex variables must be performed. It can be done using any symbolic manipulation software. We omit the details of this step and present in the next Sub-section only the result.

### B. Parameters of the 2D-PDF

We remain in the domain of Gaussian values and there is no surprise that $V$, $W$ and the geometric parameters of the cloud are some functions of means and variances of two initial random processes, i.e. of $\tau, \tau^2$ and $\sigma^2$. To present these functions in a compact form we introduce four simple expressions:

$$\begin{aligned} f_1 &= (1+\tau^2)^2, \\ f_2 &= 1+4\tau^2, \\ f_3 &= 1-\tau^2, \\ f_4 &= 1+3\tau^2. \end{aligned} \quad (43)$$

All $f_i$ equal one when $\tau = 0$, i.e. when the exponentially distributed second process is absent and phase is a Gaussian variable with zero mean.

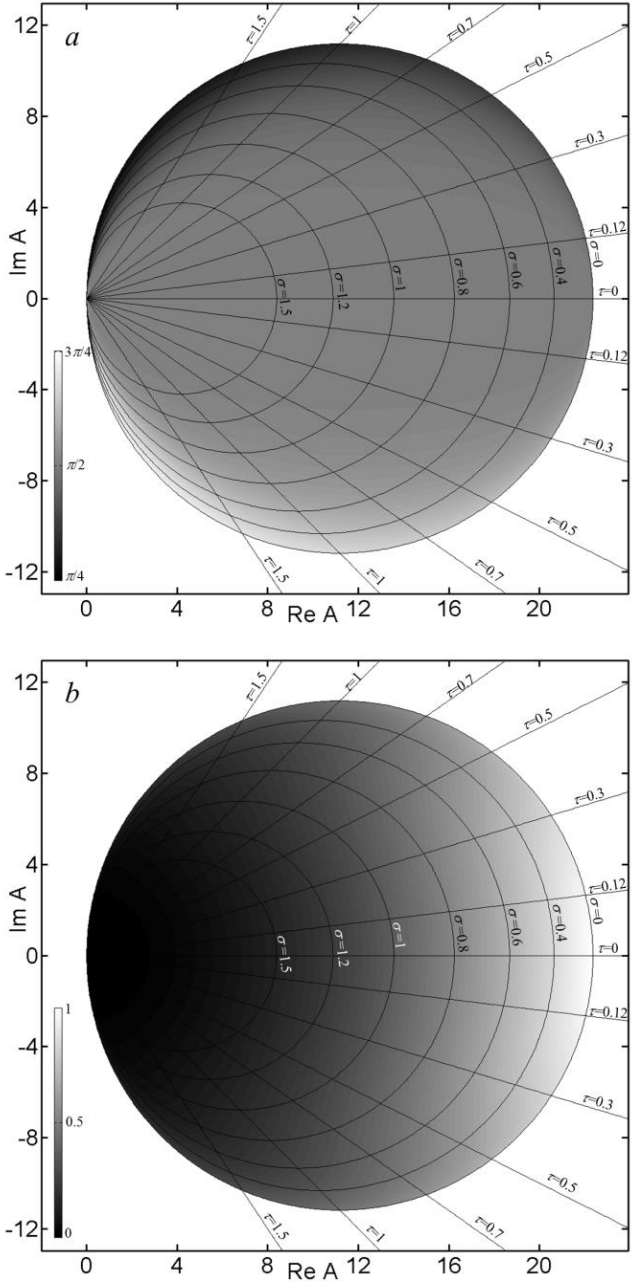

Fig. 4. Inclination angle (a) and eccentricity (b) as functions of ellipse center: $(\tau, \sigma)$ – mapping. The number of phasor $n = 500$.

First, we study the location of the center and the size of the cloud. For this we need to find $V$ and $\theta_V$:

$$\begin{aligned} V &= f_1^{-\frac{1}{4}} \exp\left(-\frac{\sigma^2}{2}\right), \\ \theta_V &= \operatorname{atan}(\varepsilon\tau). \end{aligned} \quad (44)$$

Note, that $f_1^{-\frac{1}{4}} = (1+\tau^2)^{-\frac{1}{2}} = (1+\tan^2\theta_V)^{-\frac{1}{2}} = \cos\theta_V$ and $V = \exp(-\sigma^2/2)\cos\theta_V$. To find the cloud center position $(\rho, \theta_\rho)$ we substitute Eq.(44) into Eq.(26).

When $\tau = 0$ the angular position of the center is zero. The cloud is centered on the real axis at the distance $\rho_0 = \sqrt{n}\exp(-\sigma^2/2)$ from the center of coordinates. For small $\tau$, when the contribution of the second process is weak, the angular position equals to the phase mean value: $\theta_\rho \approx \varepsilon\tau = \langle\varphi\rangle$.

When $\varepsilon\tau$ changes but $\sigma$ remains constant, the center of the cloud moves along a circular line, connecting the point $(\rho_0, 0)$ and the center of coordinates(Figure 3): anticlockwise if the P-PDF is positively skewed ($\varepsilon = 1$) or clockwise if the P-PDF is negatively skewed ($\varepsilon = -1$). The circular shape of the line can be deduced from Eq. (26) and Eq. (44): $\rho = \rho_0 \cos\theta_\rho$. When, on the contrary, $\sigma$ increases but $\varepsilon\tau$ remains constant, the center of the cloud moves along a ray making angle $\theta_\rho$ with the abscissa inwards.

While moving along a ray or a circle towards the center of coordinates the cloud expands, because increasing $\tau$ or $\sigma$ means that the P-PDF widens approaching a uniform distribution. The spread of the cloud is characterized by the total size which in this example is

$$\begin{aligned} \Sigma &= 1 - V^2 = 1 - \exp(-\sigma^2)\cos^2\theta_V \\ &\approx 1 - \exp(-\sigma^2) + \tau^2\exp(-\sigma^2). \end{aligned} \quad (45)$$

The last approximation is given for small $\tau$.

Now we study orientation of the ellipse and its eccentricity. For this we need to calculate $W$ and $\theta_W$. The real and the imaginary parts of $\boldsymbol{W}$ are

$$\begin{aligned} \operatorname{Re}\boldsymbol{W} &= \exp(-\sigma^2)\frac{f_1\exp(-\sigma^2) - f_2 f_3}{f_1 f_2}, \\ \operatorname{Im}\boldsymbol{W} &= 2\varepsilon\tau\exp(-\sigma^2)\frac{f_1\exp(-\sigma^2) - f_2}{f_1 f_2}. \end{aligned} \quad (46)$$

When $\tau = 0$ the imaginary part is zero, but the real part is negative and therefore the angle $\theta_W$ is $\pi$. If $\varepsilon = 1$ and $\tau$ is sufficiently small both, Re $\boldsymbol{W}$ and Im $\boldsymbol{W}$, are small negative values. In order to be consistent with the definition of angles presented in Figure 1 and to avoid uncertainty of atan(...) we define a piecewise function:

$$\theta_W = \begin{cases} \pi - \operatorname{atan} T_W, & \tau \leq \tau_0, \\ -\operatorname{atan} T_W, & \tau > \tau_0, \varepsilon = 1, \\ -\operatorname{atan} T_W + 2\pi, & \tau > \tau_0, \varepsilon = -1, \end{cases} \quad (47)$$

where

$$\theta_T = \frac{\operatorname{Im}\boldsymbol{W}}{\operatorname{Re}\boldsymbol{W}} = 2\varepsilon\tau\frac{f_1\exp(-\sigma^2) - f_2}{f_1\exp(-\sigma^2) - f_2 f_3} \quad (48)$$

and $\tau_0$ is a point when Re $\boldsymbol{W} = 0$, i.e. a positive root of quadratic with respect to $\tau^2$ equation: $f_1\exp(-\sigma^2) - f_2 f_3 = 0$.

In Figure 3 an arrow indicated the orientation of an ellipse. There exists certain combination $(\tau, \sigma)$ for which the cloud is oriented horizontally. This combination satisfies the condition $\theta_W = 0$ or $\theta_T = \theta_V$ for the upper half-plane and $\theta_W = 2\pi$ or $\theta_T = \pi - \theta_V$ for the lower half-plane. The combination satisfies the condition when Im $\boldsymbol{W} = 0$, and can be found as the positive root of quadratic, with respect to $\tau^2$ equation: $f_1\exp(-\sigma^2) - f_2 = 0$.

The inclination angle $\theta_T$ is calculated according to Eq. (32). When the contribution of the second process is weak ($\tau$ is small) the inclination angle can be estimated by the following approximation:

$$\theta_T \approx \frac{\pi}{2} + \frac{\varepsilon\tau^3}{\exp(-\sigma^2) - 1} \qquad (49)$$

When $\tau = 0$, the cloud is vertical ($\theta_T = \pi/2$). The deviation of the inclination angle from $\pi/2$ is proportional to $\varepsilon\tau^3$. In this term we recognize the third moment of $\varphi$ (Eq. (40)). Although the equation above cannot serve as a mathematical proof of connection between the deviation of the inclination angle from $\pi/2$ and an asymmetry of the P-PDF, it gives a feeling of this existing connection. In Figure 3, $\theta_T$ is the angle between an arrow and a ray.

Eccentricity is computed according to Eq.(30) and after some algebraic manipulations becomes:

$$e = \exp(-\sigma^2)\frac{\sqrt{f_1 \exp(-2\sigma^2) - 2f_4 \exp(-\sigma^2) + f_2}}{\sqrt{f_1 f_2} - \sqrt{f_2 \exp(-\sigma^2)}} \qquad (50)$$

In absence of the second process, i.e. when $\tau = 0$ and $\theta_\rho = 0$ the eccentricity equals $e = \exp(-\sigma^2) = 1 - \Sigma$: the vertical cloud with its center on the abscissa has a maximal elongation and a minimal spread. For small $\tau$ the following approximation is valid: $e \approx \exp(-\sigma^2) - \tau\exp(-\sigma^2)$, and the relation $e = 1 - \Sigma$ is preserved. With an increase of $\tau$ it is not anymore the case.

Figure 3 shows the set of 2D-PDFs plotted with use of Eq.(36) on a grid of constant $\tau$ (rays) versus constant $\sigma$ (circles). Crossing point between a ray and a circle is a center of an ellipse. The center can be located only within the largest circle with diameter $\sqrt{n}$ (boundary circle, $\sigma = 0$). Every point on the $(R, I)$-plane within the boundary circle is a center of an ellipse and uniquely corresponds to cetain pare $(\tau, \sigma)$. So we speak about $(\tau, \sigma)$-mapping:

$$R = \sqrt{n}\,\exp\left(-\frac{\sigma^2}{2}\right)\frac{1}{1+\tau^2},$$
$$I = \sqrt{n}\,\exp\left(-\frac{\sigma^2}{2}\right)\frac{\varepsilon\tau}{1+\tau^2}. \qquad (51)$$

The pare $(\tau, \sigma)$ gives a values for ellipse parameter. This allows plotting this parameter on the $(R, I)$-plane as a two-dimensional function. Figure 4 shows the inclinations angle and the eccentricity in this representation. Not only the geometric parameters of ellipse but also all values derived from them can be presented on the $(R, I)$-plane through $(\tau, \sigma)$-mapping. Figure 6 in Section 4 illustrates it on example of intensity standard deviation.

### C. Numerical simulation

We simulate the random phasor sum with MATLAB R2014. A phase value is generated as a sum of two random numbers: the first one with the normal distribution $N(0, \sigma)$ and the second one with the exponential distribution $E(\tau)$. The phase is used to generate a complex phasor with unit length and the random phasor sum contains 500 independent phasors. The case $\varepsilon = -1$ is simulated by taking the phase value as a difference between $N(0, \sigma)$-distributed number and $E(\tau)$-distributed number. Every outcome is plotted as a point on the $(R, I)$-plane. For every combination $(\tau, \sigma)$ up to 600 points are generated to create a cloud. The smaller clouds are created with the smaller number of points. We have chosen this simple way to generate the complex amplitude and not the full Fourier optics propagation to avoid complications related to generation of rough surface with given statistics. In other words, we deal here only with the second part of the optical task discussed in Section 2A.

Figure 5 shows the simulated clouds when $\sigma$ is fixed to 0.2 (the statistics of the first random value does not change), while $\tau$ increases from 0 to 10. To compare with the theory we plotted an analytical ellipse given by Eq.(36) for every $\tau$ and the circle $\rho = 21.9\cos\theta_\rho$. The behavior is as it is described above: when $\tau$ increases the ellipse moves along the circle towards zero, widens, rounds and rotates. It starts from the vertical orientation and rotates slower than the tangential to the circle, therefore it assumes the horizontal orientation not at $\tau = 1$, but later at $\tau_h = 1.48$.

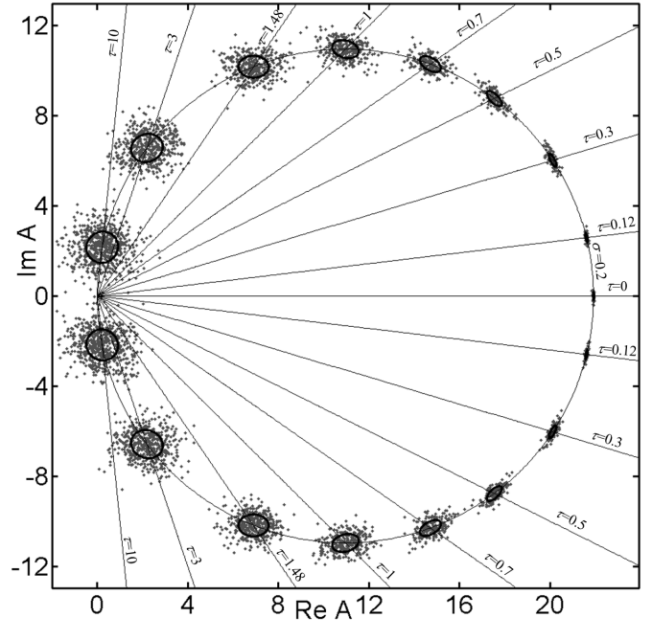

Fig. 5. Two-dimensional PDF of complex amplitude for exponentially modified normal phase distribution. The number of phasors equals 500; $\sigma$ equals 0.2 rad. The theoretical ellipses are plotted in solid lines. The results of the simulation for the same set of parameters are shown in dots.

## 4. STATISTICS OF INTENSITY

The main subject of this paper is joint distribution of the real and the imaginary parts of the complex amplitude, which we refer to as probability cloud or 2D-PDF. Although we wish to keep the focus on the geometry of the cloud itself, it is important to show how the knowledge of geometry helps to predict the statistical behavior of the measurable quantity, i.e. intensity. The standard procedure is as following. First step is finding the joint distribution of modulus and angle of the complex amplitude by moving to the polar coordinates $(A, \theta_A)$ from the mutual distribution of the real and the imaginary parts. Second step is passing to a marginal statistics $p_A(A)$ by integrating over $\theta_A$. The final step is obtaining a distribution of intensity by substituting $= \sqrt{I_s}$. (We use notation $I_s$ to distinguish intensity from the imaginary part of the complex amplitude.) Except for a small number of cases, the integration over $\theta_A$ is either insoluble analytically or the result is complicated and inapplicable. For example, Equation 9 from [4] gives the probability of intensity through an infinite sum of products modified Bessel functions; and the sum is multiplied by an exponential function. It is a very neat formula, but to our knowledge no mathematical software can process it.

The representation of the cloud through its geometric parameters allows obtaining the statistics of intensity without passing through the painful integration over the angle. To avoid overloading this paper, we give only some clues of the procedure leading to the result. We plan to present more details in the next paper.

In Section 2, Figure 2, we introduced the coordinates $(x', y')$. The 2D-PDF written in these coordinates

$$p(x', y') = \frac{1}{\sqrt{2\pi\sigma_x^2}} \exp\left[-\frac{(x'-\mu_x)^2}{2\sigma_x^2}\right] \frac{1}{\sqrt{2\pi\sigma_y^2}} \exp\left[-\frac{(y'-\mu_y)^2}{2\sigma_y^2}\right] \quad (52)$$

describes the join distribution of two independent normally distributed variables, $x'$ and $y'$ with the means $\mu_x = \rho \cos\theta_T$ and $\mu_y = \rho \sin\theta_T$ and the variances $\sigma_x^2$ and $\sigma_y^2$. Eq.(52) is obtained by recognizing that Eq.(36) and Eq.(12) describe the same curve and that determinant $D_c$ is invariant to any linear transformation of coordinates. These variables $x'$ and $y'$ are related to the intensity as

$$x' = \sqrt{I_s}\cos\theta', \quad (53)$$
$$y' = \sqrt{I_s}\sin\theta'.$$

The intensity is a sum of squares of two independent normally distributed random variables with known means and variances:

$$I_s = x'^2 + y'^2. \quad (54)$$

We do not need to integrate over $\theta'$.

Now we can follow the standard methods of statistics to obtain a moment generating function for intensity. We postpone the details of this calculation until the next paper and present only the outcome:

$$\Xi(t) = \frac{1}{\sqrt{1 - 2t + t^2(1 - e^2)}} \quad (55)$$
$$\times \exp\left[\frac{\rho^2 t}{\Sigma} \cdot \frac{1 - t(1 - e\cos 2\theta_T)}{1 - 2t + t^2(1 - e^2)}\right].$$

Due to the rotational symmetry (intensity is insensitive to the mean phase) the angular position of ellipse $\theta_\rho$ is not present in this result. The moments of intensity are the partial derivatives of this function:

$$\langle I^n \rangle = \Sigma^n \left.\frac{\partial^n \Xi(t)}{\partial t^n}\right|_{t=0}. \quad (56)$$

and yield in polynomial expressions of $\rho, \Sigma, e$ and $\cos 2\theta_T$. These values, as we have studied in the paper, are linked to statistics of phase.

The central moments $\mathcal{M}_I^n = \langle (I - \langle I \rangle)^n \rangle$ are therefore also some polynomials of the same values. We present here the expressions for the first three:

$$\mathcal{M}_I^1 = \rho^2 + \Sigma,$$
$$\mathcal{M}_I^2 = \Sigma[2\rho^2(1 + e\cos 2\theta_T) + \Sigma(1 + e^2)], \quad (57)$$
$$\mathcal{M}_I^3 = 2\Sigma^2[3\rho^2(1 + 2e\cos 2\theta_T + e^2) + \Sigma(1 + 3e^2)].$$

Figure 6 shows intensity standard deviation $\sigma_s = \sqrt{\mathcal{M}_I^2}$ for exponentially modified normal P-PDF from the previous section through $(\tau, \sigma)$ -mapping. This function has two local minima: the first one is located at the center of coordinates and the second one – at the opposite extreme of the boundary circle. The first minimum ($\sigma_s = 1$) corresponds to the case studied by Lord Rayleigh of well-developed speckles. The second minimum ($\sigma_s = 0$) corresponds to the situation (also discussed in his paper) when there is no randomness whatsoever and all phasors are identical. Between these two minima there is a ridge along which $\sigma_s$ is maximal.

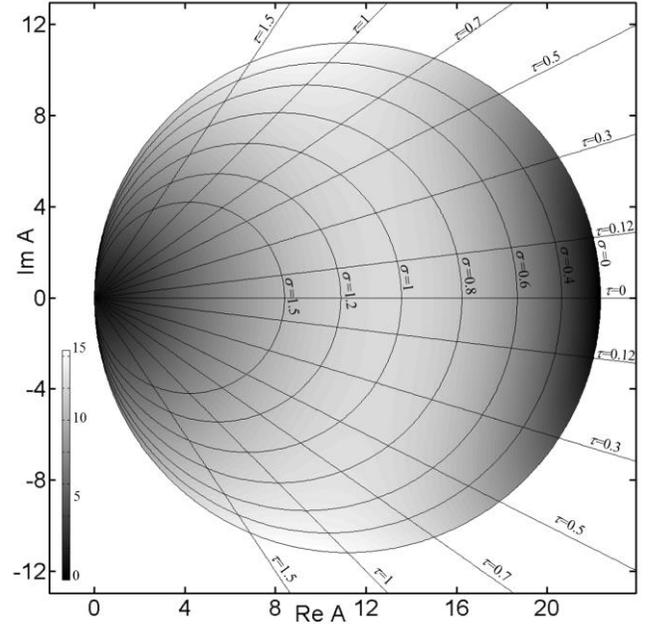

Fig. 6. Intensity standard deviation through $(\tau, \sigma)$-mapping. The number of phasors $n = 500$.

## 5. CONCLUSION

Mutual distribution of the real and the imaginary parts of complex amplitude – the resultant of a large number of vibrations, is best visualized as a probability cloud. The cloud is a powerful tool to describe statistics of light with phase fluctuation. The geometric parameters of the cloud: its position, extension, elongation and orientation, – are directly related to statistics of phase and involve only two values of phase characteristic function (which in general are complex values). There are common rules applied to the behavior of these parameters independently of a particular phase distribution: the cloud always widens when its center approaches the center of coordinates; deviation of the inclination angle from $\pi/2$ indicates certain asymmetry in initial phase distribution; eccentricity tends to one when the cloud shrinks into a point. The demonstration of the last rule (not presented in the paper) involves expanding $V$ in series at $\Sigma = 0$.

We illustrate the method of geometrical description on example of exponentially modified normal distribution and introduce the principle of $(\tau, \sigma)$ -mapping. This principle is based on the fact that there is a one-to-one correspondence between the vector $(\tau, \sigma)$ and a vector pointing to ellipse center. If for another phase distribution there is no unique relation between parameters of the P-PDF and a point on the complex plane, it nevertheless can be found by combining the parameters. In the considered example of exponentially modified normal distribution we assumed that the mean of normally distributed component is zero. If it is not zero but equals certain $\mu$, the mapping is still possible through combined parameter $\tau' = \varepsilon\tau + \mu$.

The geometrical representation of the two-dimensional probability density function of complex amplitude allows for obtaining the moment-generating function of intensity in a closed form and hence expressions for all statistical moments of intensity through the parameters of ellipse. All central moments of intensity are positive regardless of the statistics of phase. Nevertheless, simulations and experimental data [16] give negative values for some odd moments of intensity or the values based on these moments (skewness, for example). This discrepancy is not due to an error in our calculation, but to limitations of the postulate lying in its base: assuming a large number of phasor, applying the central limit theorem and writing the 2D-PDF as two-dimensional Gaussian function we enforce an elliptical shape of the cloud. The odd moments of intensity are sensitive to cloud non-ellipticity, i.e. to the number of phasors, and therefore cannot be estimated within the frame of the central limit theorem.

Continuation of this work has three directions. First task is to relate the number of phasors ($n$) to the second-order statistics (autocorrelation function) of phase. Our preliminary calculations show that assigning diameter of a zone to a coherence length yields in unrealistically small $n$. However, this result requires a more thorough investigation. Second task is to unfold Section 4 of the present paper: having obtained the general expression for moment generating function we must study this expression in detail. In particular, we shall see if it gives the correct result for the known intensity statistics included in it: exponential decay, Rician and modified Rayleigh distributions. Third task is to explore the domain of non-Gaussian complex random variables and try to find not only expressions for intensity moments by means of combinatorics (these expressions have been obtained in [17]), but the shape of the 2D-PDF. One way to do this is to assume $n$ to be large but not infinite and retain the second term in Euler's theorem.

**Acknowledgement.**

I wish to thank my colleagues: Szymon Gładysz from Fraunhofer Institute for Optronics and Jérôme Paufique from European Organisation for Astronomical Research in the Southern Hemisphere, whose constructive remarks and corrections helped improving the quality of the paper.